\documentclass{article}
\usepackage{amsmath}

\usepackage{graphicx}

\begin{document}

\title{Delay, resonance and the Lambert W function}


\author{Kenta Ohira$^{1}$ and Toru Ohira$^{2}$\\
\small \ $^{1}$Future Value Creation Research Center,\\ 
\small Graduate School of Informatics, Nagoya University, Japan\\
\small\ $^{2}$Graduate School of Mathematics, Nagoya University, Japan
}

\maketitle

\begin{abstract}%
We discuss a new type of delay differential equation that exhibits resonating transient oscillations. The power spectrum peak of the dynamical trajectory reaches its maximum height when the delay is suitably tuned. Furthermore, our analysis of the resonant conditions for this equation has revealed a new connection between the solutions of the transcendental trigonometric equation and the Lambert $W$ function. These results offer fresh insights into the nonlinear dynamics induced by delayed feedback.
\end{abstract}


\section{Introduction}

There has been interest in investigating the effect of delays in various fields such as biology, mathematics, economics, and engineering \cite{heiden1979,bellman1963,cabrera_1,hayes1950,insperger,kcuhler,longtinmilton1989a,mackeyglass1977,miltonetal2009b,ohirayamane2000,smith2010,stepan1989,stepaninsperger,szydlowski2010}). 
Typically, delays introduce oscillations and complex behaviors to otherwise simple and well-behaved systems. Longer delays are known to induce an increase in the complexity of dynamics. The Mackey--Glass equation\cite{mackeyglass1977}, which exhibits various types of dynamics including chaos, serves as a representative example.

Mathematical analysis of such delay systems has posed challenges. Although our understanding of delay systems has gradually improved (e.g.\cite{taylor}), further investigations and explorations are necessary.
``Delay Differential Equations'' are primarily employed and studied. One recent analytical approach to simple delay differential equations involves the use of the Lambert $W$ function\cite{corless}, which enables the expression of both formal and approximate solutions of the equation\cite{shinozaki,pusenjak2017,kentaohira2023}.

In this paper, we follow this path and analyze a simple delay differential equation using the $W$ function. Specifically, we examine the transient oscillatory behaviors of a delay differential equation that displays resonant behavior\cite{kentaohira2021}.  In this resonance, the power spectrum peak of the dynamical trajectory reaches its maximum height when the delay is suitably tuned. Our focus lies in the appearance of the power spectrum peaks, which we found to be related to a transcendental trigonometric equation. Interestingly, this condition can also be expressed in terms of the $W$ function. We establish an analytical and numerical connection between these two approaches, providing a novel utility for the $W$ function.

\section{Delay Differential Equation}

Recently, we proposed and studied the following delay differential equation\cite{kentaohira2021}:
\begin{equation}
{dX(t)\over dt} + a t X(t) = b X(t-\tau)
\label{dr}
\end{equation}
where $a \geq 0$, $b \geq 0$, $\tau \geq 0$ are real parameters, with $\tau$ interpreted as the delay. This equation is a slight extension of the well-studied Hayes equation\cite{hayes1950},
\begin{equation}
{dX(t)\over dt} + \alpha X(t) = \beta X(t-\tau)
\label{hayes}
\end{equation}
where $\alpha$ and $\beta$ are real constants.

Despite the small apparent change from equation (\ref{hayes}) to equation (\ref{dr}), with only the second term replaced by a linear function of time, their behaviors differ significantly. Particularly for equation (\ref{dr}), we have demonstrated that oscillatory transient dynamics appear and disappear as the delay increases while maintaining asymptotic stability at $X=0$.

\subsection{Analysis}

Let us first review some properties of equation (\ref{dr}) when $b=0$. With the initial condition $X(t=0) = X_0$, the solution to the equation is as follows:
\begin{equation}
X(t) = X_0 e^{- {1\over 2}a t^2}
\label{sho}
\end{equation}
Thus, this solution exhibits a Gaussian shape. In this case, we also note that equation (\ref{dr})  represents the ground state of the quantum simple harmonic oscillator, with 
$t$ interpreted as a position rather than time (e.g.\cite{sakurai}).

The case where $a=0$ is a special case of equation (\ref{hayes}). In this scenario, the origin $X=0$ is asymptotically stable only within the range of
\begin{equation}
- {\pi/{2 \tau} }< b <0.
\label{hayec}
\end{equation}

For the general case with $a>0, b>0$ with the delay $\tau = 0$, the solution for $X(t=0) = X_0$ is obtained as:
\begin{equation}
X(t) = X_0 e^{- {1\over 2}a t^2 + b t}
\label{tau0}
\end{equation}
This solution also exhibits a Gaussian trajectory with its peak at $b/a$.

For the case with $a>0, b>0$ and the delay $\tau \rightarrow \infty$, the dynamics are influenced by the initial function for all $0 \leq t$. Thus, for the initial function  $X(t) = X_0, (-\tau \leq t \leq 0)$,
we can replace the right-hand side of equation (\ref{dr}) as $b X(t-\tau)\rightarrow X_0$. The solution can be obtained as: 
\begin{equation}
X(t) = X_0 e^{- {1\over 2}a t^2} (1 + b \int_0^t e^{ {1\over 2}a s^2} ds) = 
X_0 e^{- {1\over 2}a t^2} (1 + b \sqrt{\pi \over {2 a}} \it{erfi}{(}\sqrt{a \over {2}} t {)}) 
\label{tauinf}
\end{equation}
where $\it{erfi}(x)$ is the imaginary error function defined as:
\begin{equation}
\it{erfi}(x) = {2\over \sqrt{\pi}} \int_0^x e^{s^2} ds
\end{equation}
The shape of this function is also a single-peaked function approaching the origin $X=0$.

Now we observe one of the major differences between equations (\ref{dr}) and (\ref{hayes}).
In the latter, the asymptotic stability of $X=0$ is lost for larger delays with $0< \alpha < \beta$, while in the former, it is maintained even with a large delay for all  $a>0, b>0$. Additionally, although both equations exhibit transient oscillations, equation (\ref{dr}) shows coherent oscillations with the tuned value of the delay $\tau$. We will now focus on these resonating phenomena.

\subsection{Power Spectrum and Resonance}

The transient dynamics of equation (\ref{dr}) are investigated through numerical simulations. Some examples are shown in Fig. 1. When the delay is 0, the dynamics have a trajectory of the Gaussian shape as shown in the previous subsection. With the increasing delay, oscillations appear in a manner such that they reside on this Gaussian dynamics. As we further increase the value of the delay, we begin to see trains of decaying pulses that have intervals of the delay time. With asymptotically large delay, the pulses disappear and the dynamics become a single peaked shape (\ref{tauinf}). On the other hand, increasing the delay does not change the asymptotic stability of $X=0$, as mentioned earlier. 

This property is in contrast to that of equation (\ref{hayes}) in which increasing the delay induces oscillations, eventually leading to the loss of the stability of $X=0$. Thus, it is different from the phenomenon called ``stability switching'' (e.g.\cite{yan})  that takes the delay as the bifurcation parameter. Coupled delay differential equations also show transient oscillations called ``delay induced transient oscillation (DITO)" \cite{milton_mmnp,pakdamanetal1998a}. However, it differs from the transient dynamics of equation (\ref{dr}) as, in DITO, the increasing delay leads to persistent and prolonged duration of oscillations.

\begin{figure}
\includegraphics[height=12cm]{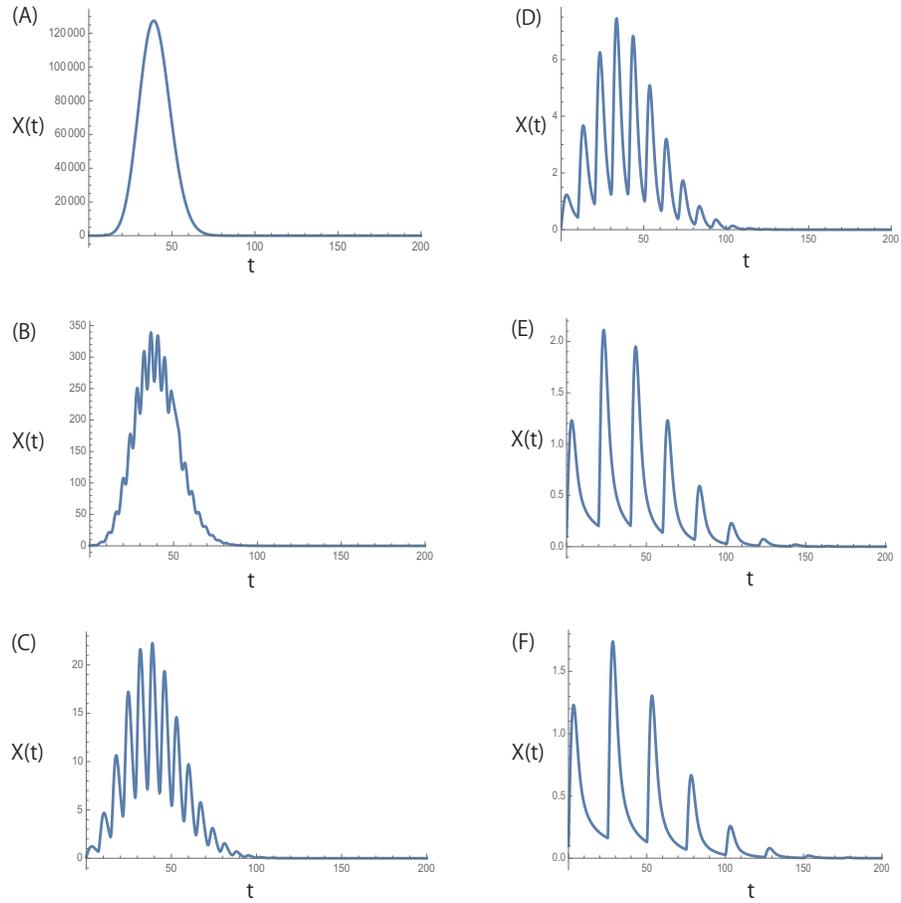}
\caption{Examples of dynamics of equation (1) for various values of the delays, $\tau$: (A)$2$, (B)$4$, (C)$7$, (D)$10$, (E)$20$, (F)$25$. The other parameters are fixed at $a=0.15, b= 6.0$ with the initial interval condition as
$X(t) = 0.1 (-\tau \leq t \leq 0)$.}
\label{dynamics}
\end{figure}

We investigate these oscillatory behaviors for the case $a > 0$, $b > 0$, and finite $\tau \neq 0$ by taking the Fourier transform of equation (\ref{dr}).
\begin{equation}
i\omega \hat{X}(\omega) + i a {d\hat{X}(\omega)\over d\omega} = - b \hat{X}(\omega) e^{i \omega \tau}
\label{ft}
\end{equation}
where
\begin{equation}
\hat{X}(\omega) = \int_{-\infty}^{\infty} e^{i \omega t} X(t) dt
\label{ft2}
\end{equation}
The solution is given as
\begin{equation}
\hat{X}(\omega) = {\cal{C}} Exp[ {- {1\over 2 a} \omega^2 + {b\over \tau a} e^{i \omega \tau}}] 
\label{ftsol}
\end{equation}
with ${\cal{C}}$ as the integration constant. We can calculate the power spectrum from equation (\ref{ftsol}).
\begin{equation}
S(\omega) = |\hat{X}(\omega)|^2 = \hat{X}(\omega)\hat{X}^*(\omega) = {\cal{C}}^2 Exp[ {- {1\over a} \omega^2 + {2 b\over \tau a} \cos{\omega \tau}}] 
\label{powersp}
\end{equation}

\begin{figure}
\includegraphics[height=12cm]{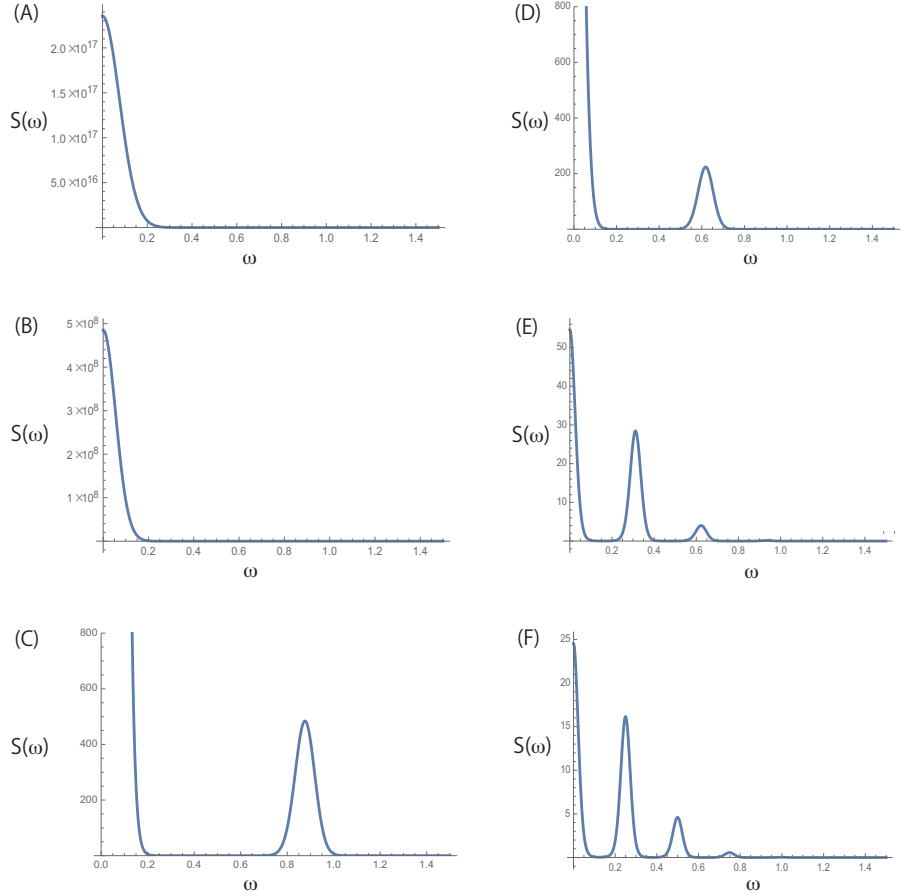}
\caption{Power spectrums given by equation (\ref{powersp}) corresponding to the dynamics shown in Fig. 1. The parameters are fixed at the same as Fig.1; $a=0.15, b= 6.0, {\cal{C}}=1$ with the initial interval condition as $X(t) = 0.1 (-\tau \leq t \leq 0)$. The delays $\tau$ are (A)$2$, (B)$4$, (C)$7$, (D)$10$, (E)$20$, (F)$25$.}
\label{power}
\end{figure}

Fig. 2 shows some representative power spectra plotted using equation (\ref{powersp}) with the same parameter setting as in Fig. 1.

In previous work, we noted that with a suitably tuned value of the delay, the peak of the power spectrum reaches its maximum height. A higher peak indicates a more coherent oscillation. It is in this sense that we observe the existence of resonance with the delay as a tuning parameter.

Also, we can further analyze this power spectrum by considering  (\ref{powersp}) as a function of both
$\omega$ and $\tau$. For example, if we define ($\omega_r, \tau_r$) as the values obtaining the maximum of the power spectrum, $Max(S)$, we can show they are given by the following:
 \begin{equation}
\tau_r = {{1\over{b}}{\sqrt{1-\delta^2}\over{\delta^2}}}, \quad \omega_r = b\delta, \quad Max(S) = S(\tau_r, \omega_r) = {\cal{C}}^2 Exp[{b^2\over a}{\delta^2}],
\label{DDE1_power_spectrum_maximum_tau_omega_with_delta0}
\end{equation}
with $\delta \approx 0.1612$ for $ a, b >0$. This result aligns with our numerical simulations. (Details of this analysis are discussed in the appendix).


\subsection{Peaks of the Power Spectrum}
Now, we focus on the analysis of these power spectrum peaks. The appearance and disappearance of the peaks in the power spectrum indicate the degree of coherence of oscillatory behaviors. The locations of the minimum and maximum points of the power spectrum function can be obtained by taking the derivative of equation (\ref{powersp}). Thus, we find that they occur at $\omega$ satisfying the equation:

\begin{equation}
\omega = - b \sin{\omega \tau}.
\label{tange}
\end{equation}

To determine the  $\omega$ values, we can examine the intersection points of the two functions from both sides of this equation. From Fig. 2, we can infer that the second non-zero smallest intersection point provides the position of the first peak (the first one corresponds to the minimum before the peak).

We can also deduce the condition for the appearance of the sequence of power spectrum peaks. Each peak emerges when these two functions are tangential to each other (see Fig. 3). This implies that not only is equation (\ref{tange}) satisfied, but also the derivatives of both sides of (\ref{tange}) are equal. Hence, we have the following conditions for the appearance of the nth peak.
\begin{equation}
\omega = - b \sin{\omega \tau},\quad 1 = - b \tau \cos{\omega\tau}, \quad {{(2n-1)\pi} \over \tau} < \omega < {{(2n-{1\over 2})\pi} \over \tau}, (n = 1,2,\dots).
\label{cond}
\end{equation}
If we set $\lambda = b \tau$,$\theta = \omega \tau$, the above condition leads to
\begin{equation}
\theta = - \lambda \sin{\theta},\quad 1 = - \lambda \cos{\theta}, \quad {(2n-1)\pi} < \theta < {(2n-{1\over 2})\pi}, (n = 1,2,\dots).
\label{cond2}
\end{equation}

From this set of equations, we can numerically estimate the solutions 
$(\theta_n, \lambda_n)$ for each $n$. First, we can derive that
$\theta_n$ and $\lambda_n$ are related by
\begin{equation}
\lambda_n^2 = \theta_n^2 +1 
\end{equation}
Then, the followings are obtained: 
\begin{equation}
\theta =  - \sqrt{\theta^2 + 1}\sin{\theta},\quad  {(2n-1)\pi} < \theta < {(2n-{1\over 2})\pi}, (n = 1,2,\dots),
\label{cond3}
\end{equation}
or
\begin{equation}
\theta =  \tan{\theta},\quad  {(2n-1)\pi} < \theta < {(2n-{1\over 2})\pi}, (n = 1,2,\dots),
\label{cond4}
\end{equation}
or
\begin{equation}
- {1 \over \lambda} = \cos{\sqrt{\lambda^2 -1}},\quad {\sqrt{{((2n-1)\pi)^2} + 1}} < \lambda < {\sqrt{{((2n-{1\over 2})\pi)^2} + 1}}.
\end{equation}
The values of the solutions $(\theta_n, \lambda_n)$ are listed in the Table \ref{trigo}. 
\begin{table}[htbp]
\begin{center}
\begin{tabular}{|c|c|c|c|} \hline
$n$ & $\theta_n$ & $\lambda_n$ & $\tan{\theta_n}$ \\  \hline
1 & 4.49341 & 4.60334 & 4.49341\\ 
2 & 10.9041 & 10.9499 & 10.9041\\
3 & 17.2208 & 17.2498 & 17.2208\\
4 & 23.5195 & 23.5407 & 23.5195\\
5 & 29.8116 & 29.8284 & 29.8116\\
6 & 36.1006 & 36.1145 & 36.1006\\
7 & 42.3879 & 42.3997 & 42.3879\\
8 & 48.6741 & 48.6844 & 48.6741\\
9 & 54.9597 & 54.9688 & 54.9597\\
10 & 61.2447 & 61.2529 & 61.2447\\ \hline
\end{tabular}
\end{center}
\caption{Numerically estimated values of $\theta_n$, $\lambda_n$ and $\tan{\theta_n}$}
\label{trigo}
\end{table}
\vspace{2em}

\begin{figure}
\begin{center}
\includegraphics[height=3.6cm]{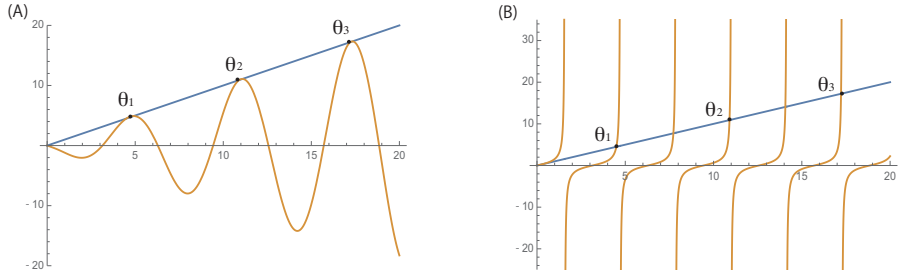}
\caption{The plots of equations (A) (\ref{cond3}) and (B) (\ref{cond4}).}
\end{center}
\label{peakcond}
\end{figure}

\subsection{Lambert $W$ function}

In the previous subsection, we discussed the solutions $(\theta_n, \lambda_n)$, and we now present an alternative method for obtaining these solutions using the Lambert $W$ function.

The Lambert $W$ function is a multivalued complex function with a complex variable $z$ that satisfies the equation:
\begin{equation}
z = W_k(z) e^{W_k(z)},
\label{wdef}
\end{equation}
where $k$ is an integer indicating the branch number\cite{corless}.
It has been noted that the $W$ function can be employed to express both formal and approximate solutions of simple delay differential equations\cite{shinozaki,pusenjak2017,kentaohira2023}. In this context, we introduce another way in which the W function can be effectively utilized.

We start with (\ref{cond2}) and use $e^{- i \theta} = \cos{\theta} - i \sin{\theta}$ to
obtain
\begin{equation}
1 - i \theta = -  \lambda e^{- i \theta}.
\label{qw}
\end{equation}
By defining $ Q \equiv - 1 + i \theta$ we can rewrite (\ref{qw}) as 
\begin{equation}
Q e^Q = {\lambda \over e}.
\label{qw2}
\end{equation}
We can now use the $W$ function on (\ref{qw2}), and $Q$ can be expressed as
\begin{equation}
Q  = {W_k}({\lambda \over e}).
\label{qw3}
\end{equation}
By the constraints that $\theta$ is real, the real part of $Q$ must be equal to $-1$, or
\begin{equation}
Re[Q]  = Re[{W_k}({\lambda \over e})] = -1.
\label{qw4}
\end{equation}
Also by the definition of $Q$, we have $\theta$ from the imaginary part,
\begin{equation}
\theta = Im[Q]  =  Im[{W_k}({\lambda \over e})].
\label{qw5}
\end{equation}
\vspace{1em}

Further, we can prove the following:
\vspace{1em}

\noindent
{\bf{Lemma 1}}
For a complex variable $z$, we have the following:
\begin{equation}
Re[{W_k}({z \over e})] = -1 \iff |{W_k}({z \over e})| = | z |
\label{wt1}
\end{equation}

\noindent
{\bf{Proof}}

\noindent
\underline{Necessary Part}:

By the definition of the $W$ function, 
\begin{equation}
{W_k}({z \over e})Exp[{W_k}({z \over e})] = {z \over e}
\label{wt2}
\end{equation}
Also, by the assumption of 
\begin{equation}
Re[{W_k}({z \over e})] = -1,
\label{wt3}
\end{equation}
we can write 
\begin{equation}
{W_k}({z \over e}) = -1 + i \mu, \quad (\mu \in R).
\label{wt4}
\end{equation}

Then, (\ref{wt2}) leads to
\begin{equation}
{z \over e} = {W_k}({z \over e})Exp[{W_k}({z \over e})] = {W_k}({z \over e})Exp[ -1 + i \mu].
\label{wt5}
\end{equation}

Thus, 
\begin{equation}
{W_k}({z \over e})Exp[ i \mu] = z, 
\label{wt6}
\end{equation}
which leads to
\begin{equation}
|{W_k}({z \over e})| = | z |.
\label{wt7}
\end{equation}

\noindent
\underline{Sufficient Part}:

By (\ref{wt2}), we have 
\begin{equation}
{| z | \over e} = |{W_k}({z \over e})Exp[{W_k}({z \over e})]| = |{W_k}({z \over e})||Exp[{W_k}({z \over e})]|.
\label{wt8}
\end{equation}
Also, by the assumption
\begin{equation}
|{W_k}({ z  \over e})| =  | z |,
\label{wtas1}
\end{equation}
this leads to
\begin{equation}
{| z | \over e} = |{z}| |Exp[{W_k}({z \over e})]|
\label{wt9}
\end{equation}
If we set  
\begin{equation}
{W_k}({z \over e}) = \eta + i \mu, \quad (\eta, \mu \in R)
\label{wt10}
\end{equation}
(\ref{wt9}) can be re-writen as
\begin{equation}
{1 \over e} = |Exp[{W_k}({z \over e})]| ={e^\eta},
\label{wt11}
\end{equation}
leading to 
\begin{equation}
\eta = Re[{W_k}({z \over e})] = -1
\label{wt12}
\end{equation}

\noindent
\underline{Q.E.D.}
\vspace{1em}

We are now in a position to put together pieces obtained through the analysis of resonant peaks. 
They can be summarized as follows.
\vspace{3em}

\noindent
{\bf{Theorem 1}}

Let $(\theta,z)$ satisfy the following,
\begin{equation}
\theta = - \lambda \sin{\theta},\quad 1 = - \lambda \cos{\theta}, \quad {(2n-1)\pi} < \theta < {(2n-{1\over 2})\pi}, (n = 1,2,\dots),
\end{equation}
then they also satisfy the following for some $k$
\begin{equation}
Re[{W_k}({\lambda \over e})] = -1, 
\end{equation}
and
\begin{equation}
\theta = Im[{W_k}({\lambda \over e})], \quad | \lambda | = |{W_k}({\lambda \over e})|.
\end{equation}
\vspace{1em}

We can also prove the inverse of the above theorem
\vspace{2em}

\noindent
{\bf{Theorem 2}}

For any  integer $k$, there exist $\hat{\lambda}_k$ such that 
\begin{equation}
Re[{W_k}({\hat{\lambda}_k \over e})] = -1, 
\end{equation}
where ${W_k}$ is the $k$-th branch of the $W$ function. With such $\hat{\lambda}_k$,
we can define
\begin{equation}
\hat{\theta}_k = Im[{W_k}({\hat{\lambda}_k \over e})].
\end{equation}
Then, $\hat{\theta_k}$ satisfies for all integer $k$
\begin{equation}
\hat{\theta}_k = - \hat{\lambda}_k \sin{\hat{\theta}_k},\quad 1 = -\hat{\lambda}_k \cos{\hat{\theta}_k}.
\end{equation}
\vspace{1em}

\noindent
{\bf{Proof}}

The existence of $\hat{\lambda}_k$ satisfying 
\begin{equation}
Re[{W_k}({\hat{\lambda}_k \over e})] = -1, 
\end{equation}
for all $k = 0, 1, 2, \dots$ is true by the known properties of the $W$ function\cite{corless}.

Then by Eq. (25) of Lemma 1, we can write $W_k({\hat{\lambda}_k \over e})$  in two ways:

\begin{equation}
W_k({\hat{\lambda}_k \over e})] = 
Re[{W_k}({\hat{\lambda}_k \over e})]  + i Im[{W_k}({\hat{\lambda}_k \over e})] = -1 + i \hat{\theta}_k 
\end{equation}

\begin{equation}
W_k({\hat{\lambda}_k \over e}) = | \hat{\lambda}_k | {e^{i\mu_k}}, \quad (\mu \in R)
\end{equation}
With these together, we obtain the following relations:
\begin{eqnarray}
\hat{\theta}_k & = & | \hat{\lambda}_k | \sin({\mu_k})\nonumber\\
-1 & = & | \hat{\lambda}_k | \cos({\mu_k}) 
\end{eqnarray}

Now, we can show the following together with the properties of the $W$ function.
(In the following we set $W_k \equiv {W_k}({\hat{\lambda}_k \over e})$.)
\begin{eqnarray}
\cos(\hat{\theta}_k) 
& = & {1\over 2}( e^{i \hat{\theta}_k } + e^{-i \hat{\theta}_k} )\nonumber \\
& = & {1\over 2}( e^{i \hat{\theta}_k } + e^{-i \hat{\theta}_k} )\nonumber \\
& = & {1\over 2}( e^{{W_k}+1} + e^{-{W_k}-1} )\nonumber \\
& = & {1\over 2}({\hat{\lambda}_k \over W_k} + {W_k \over \hat{\lambda}_k} )\nonumber \\
& = & {\hat{\lambda}_k \over |\hat{\lambda}_k |}  {1\over 2}( e^{-i\mu_k} + e^{i\mu_k} )\nonumber \\
& = & {\hat{\lambda}_k \over |\hat{\lambda}_k |}  \cos(\mu_k)
\end{eqnarray}
In a similar way, we can show
\begin{equation}
\sin(\hat{\theta}_k) = - {\hat{\lambda}_k \over |\hat{\lambda}_k |}  \sin(\mu_k)
\end{equation}

Putting the above together, we obtain the desired result.
\begin{eqnarray}
\hat{\theta}_k & = &  - \hat{\lambda}_k \sin(\hat{\theta}_k)\\
-1 & = & \hat{\lambda}_k \cos(\hat{\theta}_k) 
\end{eqnarray}
\vspace{1em}
\underline{Q.E.D.}
\vspace{1em}

Based on the above, we further want to investigate between the $n$-th root and the $n$-th branch of the $W$ function.
The following theorem is derived and confirmed with numerical estimations.

\noindent
{\bf{Theorem 3}} 

The $n$-th root $\theta_n$ of the following,
\begin{equation}
\theta_n =  \tan{\theta_n},\quad  {(2n-1)\pi} < \theta_n < {(2n-{1\over 2})\pi}, (n = 1,2,\dots),
\end{equation}
is given by the $n$-th branch of the $W$ function
\begin{equation}
\theta_n = Im[{W_n}({\lambda_n \over e})],
\end{equation}
where $\lambda_n$ satisfies
\begin{equation}
Re[{W_n}({\lambda_n \over e})] = -1.
\end{equation}

\noindent
{\bf{Proof}}

We start with Theorem 1, and let $\theta_1 < \theta_2 < \theta_3 < \dots$ and corresponding 
$\lambda_1 < \lambda_2 < \lambda_3 < \dots$  satisfy 
\begin{equation}
\theta = - \lambda \sin{\theta},\quad 1 = - \lambda \cos{\theta}, \quad {(2n-1)\pi} < \theta < {(2n-{1\over 2})\pi}, (n = 1,2,\dots),
\label{trigo2}
\end{equation}
Then, they all satisfy the following for some $k$
\begin{equation}
Re[{W_k}({\lambda \over e})] = -1, \quad \theta = Im[{W_k}({\lambda \over e})], \quad | \lambda | = |{W_k}({\lambda \over e})|.
\label{wcond}
\end{equation}
On the other hand, Theorem 2 tells us that for any $k =1, 2, \dots$, there exist 
$(\hat{\lambda}_k, \hat{\theta}_k)$ that satisfies \ref{trigo}. Also, by the properties of the $W$ function\cite{corless}, we have $\hat{\theta}_1 < \hat{\theta}_2 < \hat{\theta}_3 < \dots$ and the corresponding $\hat{\lambda}_1 < \hat{\lambda}_2 < \hat{\lambda}_3 < \dots$.

Using ${W_0}({-1 \over e}) = -1$, we have $\theta_0 = \hat{\theta}_0 = 0$. We can now use induction.
If we assume $\theta_i \leq \hat{\theta}_i$, 
then it must be $\theta_{i +1} = \hat{\theta}_{i +1}$. This is because if  $\theta_{i +1} = \hat{\theta}_{i+s}$, $(s>2)$, then $\hat{\theta}_{i+1}$ must be equal to $\theta_{t}$ with some $ t > i+1$ by Theorem 2. However, this is in contradiction to 
the fact that $\theta_1 < \theta_2 < \theta_3 < \dots$ and $\hat{\theta}_1 < \hat{\theta}_2 < \hat{\theta}_3 < \dots$.

Hence, we reached $\theta_n = \hat{\theta}_n$ and correspondingly $\lambda_n = \hat{\lambda}_n$. As (\ref{trigo2}) is
equivalent to 
\begin{equation}
\theta_n =  \tan{\theta_n},\quad  {(2n-1)\pi} < \theta_n < {(2n-{1\over 2})\pi}, (n = 1,2,\dots),
\end{equation}
the statement of the above theorem is true.
\vspace{1em}

\noindent
\underline{Q.E.D.}
\vspace{2em}

In Table \ref{wf}, we present the results of estimated related numerical values. Comparing Tables 1 and 2 supports the above theorems. (We do not discuss further here, but Theorem 3 can be extended for $n < 0$.) Thus, through the analysis of resonant peaks, we have connected the solutions of the trigonometric transcendental function with a specific value of the $n$-th branch of the $W$ function. To the author's knowledge, this relation has not been explicitly pointed out.


\begin{table}[htbp]
\begin{center}
\begin{tabular}{|c|c|c|c|} \hline
$n$ & $\lambda_n$ & ${W_n}({\lambda_n \over e})$ & $|{W_n}({\lambda_n \over e})|$\\  \hline
1  & 4.60334 & -1.0 + i 4.49341 & 4.60334\\ 
2  & 10.9499 & -1.0 + i 10.9041 & 10.9499\\
3  & 17.2498 & -1.0 + i 17.2208 & 17.2498\\
4  & 23.5407 & -1.0 + i 23.5195 & 23.5407\\
5  & 29.8284 & -1.0 + i 29.8116 & 29.8284\\
6  & 36.1145 & -1.0 + i 36.1006 & 36.1145\\
7  & 42.3997 & -1.0 + i 42.3879 & 42.3997\\
8  & 48.6844 & -1.0 + i 48.6741 & 48.6844\\
9  & 54.9688 & -1.0 + i 54.9597 & 54.9688\\
10 & 61.2529 & -1.0 + i 61.2447 & 61.2529\\ \hline
\end{tabular}
\end{center}
\caption{Numerically estimated values of $\lambda_n$, ${W_n}({\lambda_n \over e})$ and $|{W_n}({\lambda_n \over e})|$}
\label{wf}
\end{table}

\section{Discussion}
In this paper, we have presented some properties of the Lambert $W$ function through the analysis of resonant behaviors of a simple delay differential equation. The connection between the solutions of the trigonometric transcendental equation and that of the $W$ function is revealed. It remains to be explored if these properties of the W function can be utilized in a broader context.
\vspace{2em}

\noindent
{\bf Acknowledgments}

The authors would like to thank useful discussions with Prof. Hideki Ohira and members of his research group at Nagoya University. This work was supported by "Yocho-gaku" Project sponsored by Toyota Motor Corporation, JSPS Topic-Setting Program to Advance Cutting-Edge Humanities and Social Sciences Research Grant Number JPJS00122674991, JSPS KAKENHI Grant Number 19H01201, and the Research Institute for Mathematical Sciences,
an International Joint Usage/Research Center located in Kyoto University.
\clearpage

\section*{Appendix}
\large{\bf{A-1: Detailed Analysis of Power Spectrum}}\\

We present here more thorough analysis of the power spectrum. We rewrite (\ref{powersp}) explicitly as a function of
$\omega$ and $\tau$ as follows:
\begin{equation}
S(\tau,\omega) = {\cal{C}}^2 Exp[ {- {1\over a} \omega^2 + {2 b\over \tau a} \cos{\omega \tau}}] 
\end{equation}
If we set the resonant point $(\tau_r,\omega_r)$ that gives the maximum of this function, it satisfiesWe ${\partial S\over \partial\omega}(\tau_r,\omega_r)=0$ and ${\partial S\over \partial\tau}(\tau_r,\omega_r)=0$.
Thus,
\begin{equation}
\omega_r = -b\sin{\omega_r\tau_r}, \quad \omega_r\tau_r\sin{\omega_r\tau_r} = -\cos{\omega_r\tau_r}.
\label{DDE1_condition_power_spectrum_extremum}
\end{equation}
For $b >0$, these means $\sin{\omega_r\tau_r} < 0$ and $\cos{\omega_r\tau_r} > 0$, leading to
${3\over2}\pi < \omega_r\tau_r < 2\pi$.
Then,$\omega_r$ satisfies the following and uniquely determined.
\begin{equation}
\omega_r = -b\sin{\sqrt{({b\over\omega_r})^2-1}}, \quad  {b\over{\sqrt{4\pi^2+1}}} < \omega_r < {b\over{\sqrt{{9\over4}\pi^2+1}}} .
\label{DDE1_power_spectrum_maximum_omega}
\end{equation}

Also, $\tau_r$ can be expressed using $\omega_r$:
\begin{equation}
\tau_r = {{\sqrt{b^2-\omega_r^2}}\over{\omega_r^2}}
\label{DDE1_power_spectrum_maximum_tau}
\end{equation}
With these $\tau_r,\omega_r$ the power spectrum is at its maximum. Also, at this maximum the
expression describing the power spectrum can be simplified by using 
(\ref{DDE1_condition_power_spectrum_extremum}) as follows:
\begin{equation}
Max(S) = S(\tau_r, \omega_r) = {\cal{C}}^2 Exp[{\omega_r^2\over a}]
\label{DDE1_power_spectrum_maximum}
\end{equation}
If we set $\omega_r = b\delta$ for (\ref{DDE1_power_spectrum_maximum_omega}), we obtain
\begin{equation}
\delta = -\sin{\sqrt{({1\over\delta})^2-1}}, \quad  {1\over{\sqrt{4\pi^2+1}}} < \delta < {1\over{\sqrt{{9\over4}\pi^2+1}}}.
\label{DDE1_power_spectrum_maximum_delta}
\end{equation}
By numerically solving above 
$\delta \approx 0.1612$ is attained, and using it for
(\ref{DDE1_power_spectrum_maximum_tau}) and (\ref{DDE1_power_spectrum_maximum}), the resonant point can be expressed as follows:
\begin{equation}
\tau_r = {{1\over{b}}{\sqrt{1-\delta^2}\over{\delta^2}}}, \quad \omega_r = b\delta, \quad Max(S) = S(\tau_r, \omega_r) = {\cal{C}}^2 Exp[{b^2\over a}{\delta^2}]
\label{DDE1_power_spectrum_maximum_tau_omega_with_delta}
\end{equation}

Further we can have 
$\tau_r \approx 37.9664/b$, $\omega_r \approx 0.1612b$ and $Max(S) \approx Exp[0.0259b^2/a]$ by using the numerical value of $\delta \approx 0.1612$\\

\begin{figure}[h]
\begin{center}
\includegraphics[height=6cm]{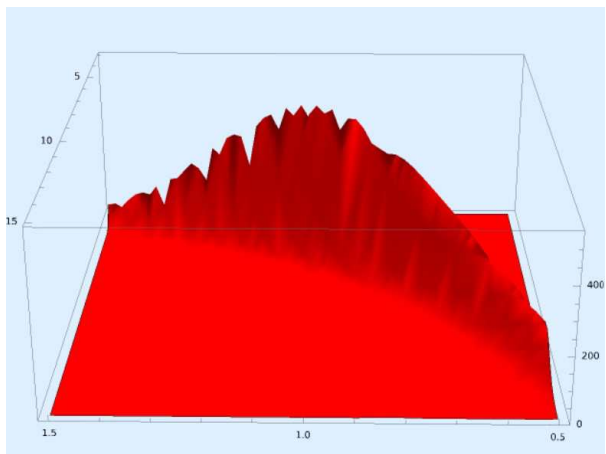}
\caption{3-D plot of the power spectrum as a function of $\omega$ and $\tau$. The parameters are set at $a=0.15$, $b= 6.0$, and the integration constant is $C=1$. The numerical maimal value estimated from this plots are given as $Max(S) \approx 512.179$ when $(\tau,\omega) \approx (6.32774, 0.967368)$.}
\label{figure_DDE1_power_spectrum_3D}
\end{center}
\end{figure}

\noindent
In order to confirm the above derivation, we examine equation (\ref{dr}) for the case with parameter values at $a=0.15$ and $b=6.0$.
For these values, equation (\ref{DDE1_power_spectrum_maximum_tau_omega_with_delta}) gives $\tau_r \approx 6.327$, $\omega_r \approx 0.9673$, $Max(S) \approx 512.179$.??We set the integration constant as $C=1$.??
Comparing these with simulations shown in Figure \ref{figure_DDE1_power_spectrum_3D}, the values of $\tau_r,\omega_r,Max(S)$ are consistent.
\vspace{1em}\\
\noindent
These results suggest that when $\tau = \tau_r$, oscillatory dynamics appears with the angular frequency $\omega = \omega_r$ as the major component.


\noindent
\large{\bf{A-2: For the case of $b < 0$}}\\

In the main text, we consider the case of $b > 0$ for equation (\ref{dr}). We can perform the same analysis for the case of $b < 0$. We observe the similar dynamical behavior of transient dynamics with power spectrum resonance. We present the representative dynamics and power spectrums in Figures. 5 and 6.
\begin{figure}
\includegraphics[height=10cm]{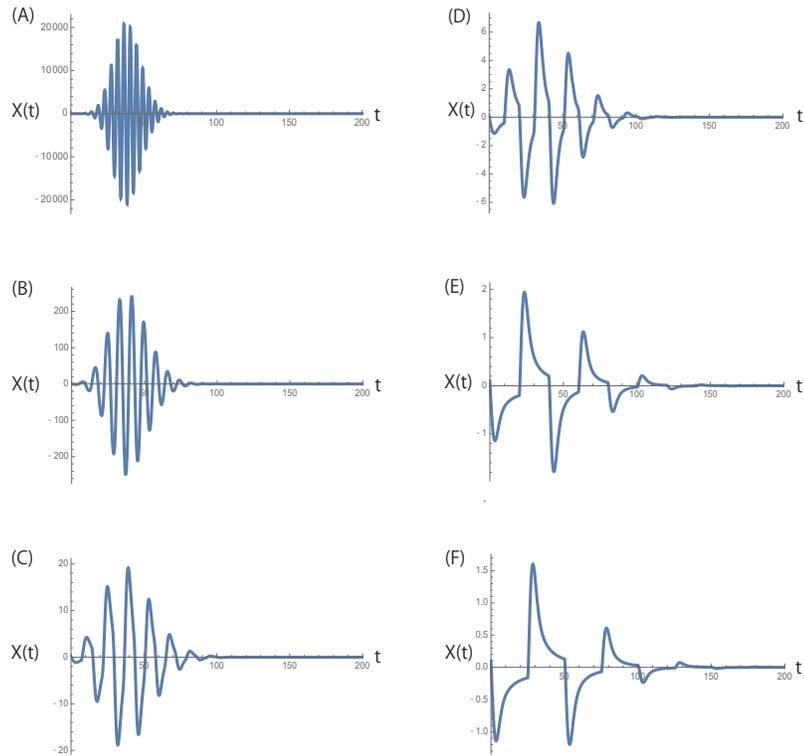}
\caption{Examples of dynamics of equation (\ref{dr}) for various values of the delays, $\tau$: (A)$2$, (B)$4$, (C)$7$, (D)$10$, (E)$20$, (F)$25$. The other parameters are fixed at $a=0.15, b= - 6.0$ with the initial interval condition as
$X(t) = 0.1 (-\tau \leq t \leq 0)$.}
\label{dynamics2a}
\end{figure}

\begin{figure}
\includegraphics[height=10cm]{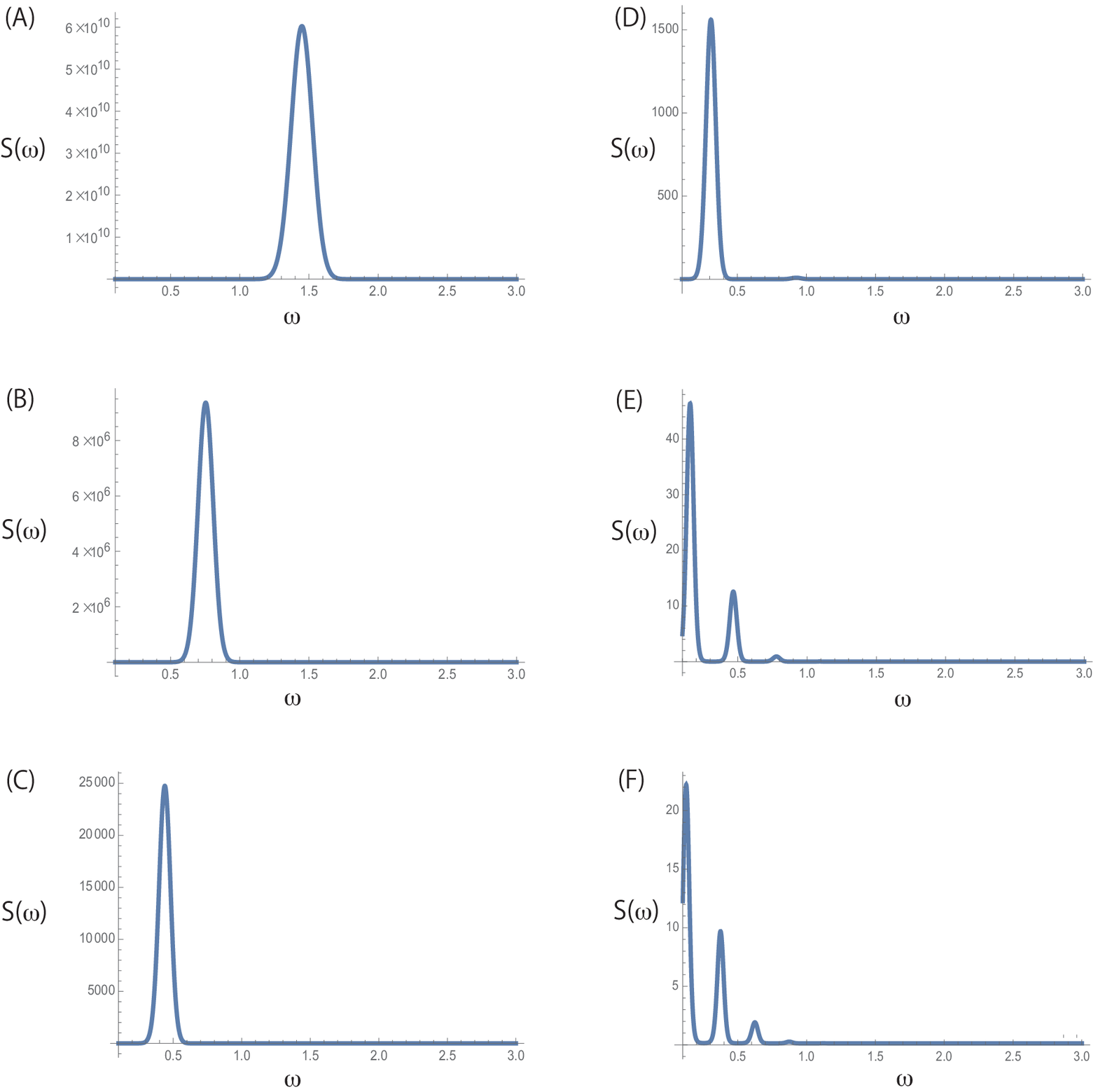}
\caption{Power spectrums given by equation (\ref{powersp}) corresponding to the dynamics shown in Fig. 5. The parameters are fixed at the same as Fig.5; $a=0.15, b= 6.0, {\cal{C}}=1$ with the initial interval condition as $X(t) = 0.1 (-\tau \leq t \leq 0)$. The delays $\tau$ are (A)$2$, (B)$4$, (C)$7$, (D)$10$, (E)$20$, (F)$25$.}
\label{power2a}
\end{figure}

The analysis of power spectrum gives that the resonance point ($\omega_r, \tau_r$) and the maximum of the power spectrum are given similarly by the following:
\begin{equation}
\tau_r = - {{1\over{b}}{\sqrt{1-\delta^2}\over{\delta^2}}}, \hspace{0.1em} \omega_r = - b\delta, \quad Max(S) = S(\tau_r, \omega_r) = {\cal{C}}^2 Exp[{b^2\over a}{\delta^2}],
\label{DDE1_power_spectrum_maximum_tau_omega_with_delta02}
\end{equation}
with $\delta \approx 0.337$.

\end{document}